\documentclass[conference]{IEEEtran}
\IEEEoverridecommandlockouts
\usepackage{cite}
\usepackage{amsmath,amssymb,amsfonts}
\usepackage{algorithmic}
\usepackage{graphicx}
\usepackage{textcomp}
\usepackage{xcolor}
\usepackage{dsfont}
\def\BibTeX{{\rm B\kern-.05em{\sc i\kern-.025em b}\kern-.08em
    T\kern-.1667em\lower.7ex\hbox{E}\kern-.125emX}}
\begin{document}

\title{Towards a Robust WiFi-based Fall Detection with Adversarial Data Augmentation}

\author{\IEEEauthorblockN{Tuan-Duy H. Nguyen}
\IEEEauthorblockA{\textit{Faculty of Information Technology} \\
\textit{University of Science, VNU-HCM}\\
Ho Chi Minh City, Vietnam \\
nhtduy@apcs.vn}
\and
\IEEEauthorblockN{Huu-Nghia H. Nguyen}
\IEEEauthorblockA{\textit{Faculty of Information Technology} \\
\textit{University of Science, VNU-HCM}\\
Ho Chi Minh City, Vietnam \\
nhhnghia@apcs.vn}
}

\maketitle

\begin{abstract}
Recent WiFi-based fall detection systems have drawn much attention due to their advantages over other sensory systems. Various implementations have achieved impressive progress in performance, thanks to machine learning and deep learning techniques. However, many of such high accuracy systems have low reliability as they fail to achieve robustness in unseen environments. To address that, this paper investigates a method of generalization through adversarial data augmentation. Our results show a slight improvement in deep learning-systems in unseen domains, though the performance is not significant.
\end{abstract}

\begin{IEEEkeywords}
channel state information (CSI), deep learning, human behavior recognition, fall detection, WiFi.
\end{IEEEkeywords}

\section{Introduction}\label{introduction}
Falls are known to be a fatal clinical problem faced by older adults \cite{rubenstein2006falls}. In addition to physical injuries, falls also cause psychological damage to elders, creating the fear of falling cycle, and consequently, lead to the functional decline in physical activities \cite{friedman2002falls}. After a fall, even without injury, elders will reduce their physical activities in responding to the fear of falling again. Such functional declines, in turn, decrease the fitness, mobility, and balance of elders, and thus, negatively affect their quality of life. To that end, timely and automatic detection of falls assisting the elders, especially those who live alone and independently, and their family is a critical need of the community \cite{10.1093/ageing/afm169}.

There have been many designs for automated fall detection systems \cite{noury2007fall, yu2008approaches}, such as \emph{wearable sensor-based systems} \cite{lord1991falls, delahoz2014survey, rashidi2013survey}, \emph{ambient device-based systems} \cite{spasova2014survey}, and \emph{vision-based systems} \cite{zhang2015survey}. Such approaches, however, suffer from inconvenience and ineffectiveness in conjunction with privacy intrusion. Recent developments in WiFi sensing has opened a new direction for fall detection \cite{ma2019wifi, wang2019survey}. WiFi systems are now essential household devices with advanced technologies like Multiple-Input Multiple-Output (MIMO) and Orthogonal Frequency-Division Multiplexing (OFDM) becoming increasingly popular in off-the-shelf commodity products. Unlike previously indicated solutions, WiFi sensing requires no attachment to the body, not intrusive, nor dependent on lighting conditions.

WiFi systems with MIMO-OFDM can provide Channel State Information (CSI), which records amplitude and phase variations of wireless signals propagated from the transmitter to the receiver. A time series of CSI captures the travel environment of signals through objects and humans in time, frequency, and spatial domains. Hence, the rationale behind WiFi sensing systems is that different human activities correspond to different patterns in the change of CSI.

However, WiFi amplitude attenuation and phase shifts are very sensitive to many different factors, such as network settings, spatial areas, objects, or humans. Most WiFi sensing systems are trained and tested in the same environments with the same actors, and many of them cannot generalize to new people in new contexts. This lack of robustness leads to low reliability in systems, especially high-capacity deep learning-based systems, as they tend to degrade significantly in unseen testing domains.
However, it is infeasible in practice to have all users and environments data trained beforehand. Thus, in this research, we explore if adapting a domain generalization method can achieve robustness for WiFi sensing systems across different testing domains. Specifically, we apply the Domain Generalization via Adversarial Data Augmentation (ADA) method \cite{domain-generalization-volpi-2018} on the FallDeFi dataset \cite{palipana2018falldefi}. Our preliminary results show that ADA contributes to a slight improvement in experimented deep learning-based methods. We plan to evaluate ADA more extensively on other deep networks, as well as on hybrid techniques for CSI processing in future works.

\section{Background and Related Works}\label{background-related}
\subsection{Channel State Information}\label{literature}
\begin{figure}
    \centering
    \includegraphics*[width=0.8\columnwidth]{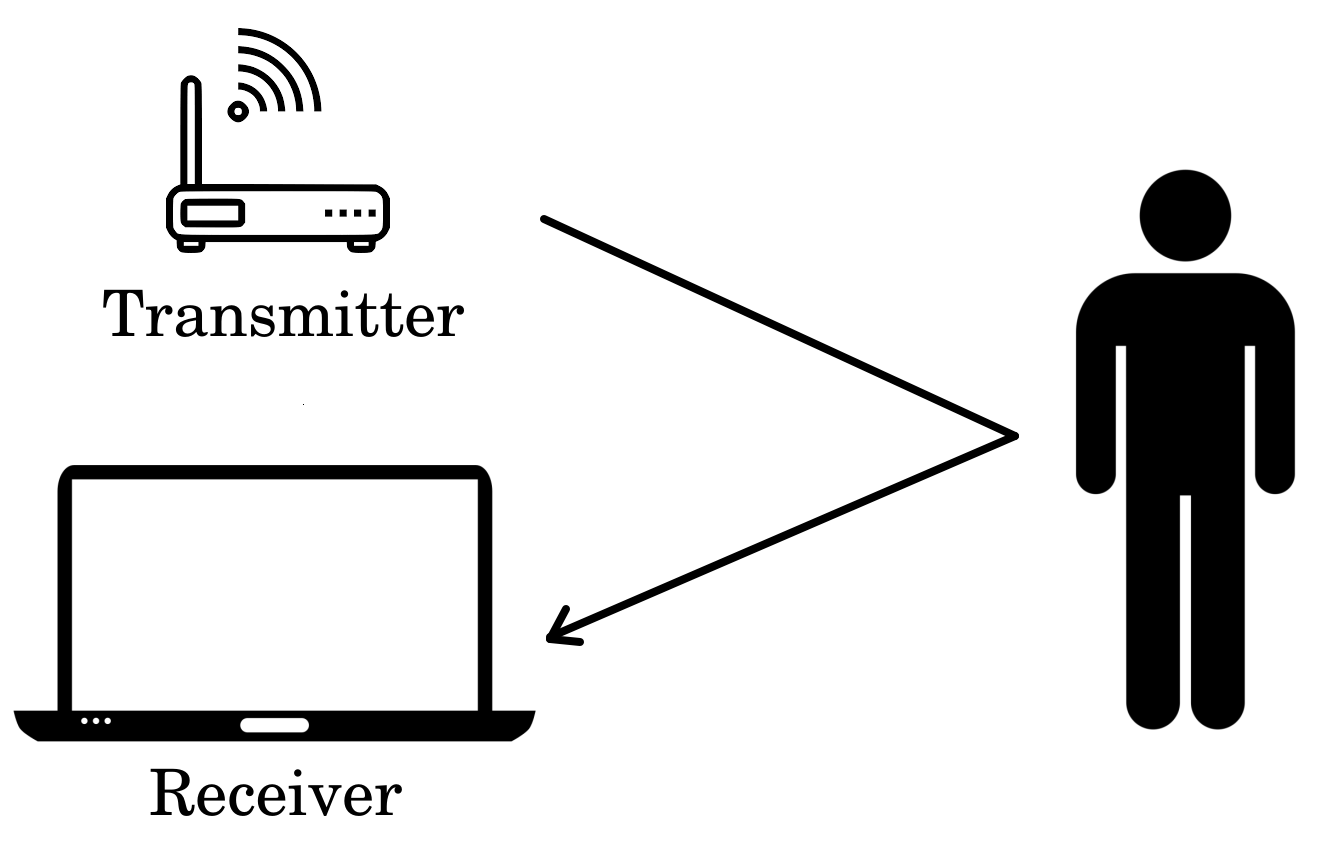}
    \caption{ WiFi interference can be caused by human activities}
    \label{fig:wave-interference}
\end{figure}

In a MIMO wireless configuration, a typical CSI matrix (or transmission matrix) is formed by transmitting a symbol from each of the transmitting antennas, and its response on the multiple receiving antennas are noted. As the physical space constraints, the propagation of wireless signals, the received signals, in turn, contain information that characterizes the environment they pass through. If a person presents in the environment, additional signal paths are introduced by the scattering of the human body (Fig. \ref{fig:wave-interference}). 

In MIMO, the system configuration typically contains M antennas at the transmitter and N antennas at the receiver. The channel response is then expressed as a transmission matrix $H$ of dimension $N \times M$. The matrix contains elements that are complex and describe both the amplitude and phase variations of the link. The direct path formed between receiving antenna $i$ and transmitting antenna $j$ is expressed as $h_{ij}$.

The received vector $y$ is expressed in terms of the channel transmission matrix $H$, the input vector $x$ and noise vector $n$ as 
\begin{equation}
    y = Hx+n
\end{equation}

Current WiFi standards (e.g., IEEE 802.11n/ac) use orthogonal frequency division modulation (OFDM) in their physical
layer. OFDM splits its spectrum band into multiple frequency sub-bands, called subcarriers, and sends the
digital bits through these subcarriers in parallel. The CSI reveals
a set of channel measurements depicting the amplitude and
phase of every OFDM subcarrier. The CSI of a single subcarrier
is in the following mathematical format $H_i = |H_i|e^{j\angle H_i}$ where $|H_i|$ and $\angle H_i$ denote the amplitude and the phase of the CSI at the $i^{th}$ subcarrier respectively.

Most of the works in WiFi sensing employ either the Intel 5300 WiFi NIC card with the corresponding CSI tool \cite{halperin2011tool} or the Atheros WiFI NIC 9k series with the Atheros CSI Tool \cite{Xie:2015:PPD:2789168.2790124}. These are used to record signals from COTS WiFi devices with multiple antennas for MIMO communication and at the physical layer adopt OFDM that supports IEEE 802.11/ac standard. In comparing, the Intel 5300 CIS tool only provides CSI for 30 subcarriers (out of the total 56 subcarriers for 20 MHz bandwidths and 114 subcarriers for 40 MHz). This limit by is addressed by the Atheros CSI Tool, which provides CSI data on all 114 subcarriers for 40 MHz bandwidth.

\subsection{Related Works}\label{related}
Various techniques have been proposed and studied to address the need for an automatic fall detection system \cite{noury2007fall,yu2008approaches}. \emph{Wearable sensor-based approaches} are among the first since the proposal Lord and Colvin for an accelerometer-based approach in 1991  \cite{lord1991falls}. Following the work are methods that employ numerous kinds of sensors, ranging from gyroscopes, barometric pressure sensors, RFID, to smartphones with combinations of sensory information\cite{delahoz2014survey,rashidi2013survey}. However, these systems only work when carried by users while the always-on-body requirement \cite{Wu:2016:WWD:2971648.2971658} for the elders is not always applicable\cite{steele2009elderly}. 

\emph{Ambient device-based approaches} try to make use of ambient information caused by falls to detect the accident. The ambient information being used includes audio noise, floor vibration, and infrared sensing data.\cite{spasova2014survey} \emph{Computer vision-based approaches} use cameras to monitor the environment, either by capture images or video sequences\cite{zhang2015survey}. However, these approaches either come with an on-body requirement, which is ineffective in many situations, a high rate of false alarm due to ambient noise, or privacy intrusion issues. 

Among Wi-Fi-based methods, one of the most popular approaches is WiFall by Han et al\cite{wang2017wifall}. Their process consists of the following three phases: Data Processing, Anomaly Detection, and Activity Classification. Firstly, in the Data Processing phase, the authors perform weighted moving average method on CSI in order to reduce environmental noise. In the Anomaly Detection phase, static human bodies create no turbulence in the CSI domain and thus create no effect in CSI, while activities such as walking, running, falling are can generate a variance in CSI. As a result, human activities can be perceived as anomalies compared to stationary humans and recognized by an anomaly detection algorithm. After the Anomaly Detection phase, any detected action would go through the Activity Classification phase. This phase involves extracting 7-feature vectors manually engineered and putting these vectors through a binary Support Vector Machine classifier to recognize fall from other human activities. On average, WiFall manages to achieve an 87\% precision detection rate with an 18\% false alarm rate when tested on the same domain used for training. Similarly, Anti-Fall \cite{zhang2015anti} and RT-Fall \cite{wang2017rt} perform manual feature engineering techniques but on both phase and amplitude information. With the additional use of phase shifts, the performance appears to improve with Anti-Fall reach a precision of 89\% and a false alarm rate of 13\%, while  RT-Fall reaches a true positive rate of 91\% and false alarm of 8\%. One out-of-the-box approach is employed in WiSpeed \cite{zhang2018wispeed} where estimation models are built to estimate the speed and acceleration of different activities. This information is then used to detect falls, with the rationale being the acceleration and speed patterns of falling down and other activities are different. WiSpeed is able to achieve an impressive true positive rate of 95\% and zero false alarms.

Besides traditional methods using complex signal processing techniques like \cite{wang2017wifall, zhang2015anti, wang2017rt, zhang2018wispeed}, deep-learning-based approaches are also popular since deep learning models are able to extract non-linear features in the hidden representation space automatically. Consequently, these methods eliminate the need for a signal processing procedure. One such deep-learning-based approach is conducted by Yousefi et al. \cite{yousefi2017survey}. Following CARM proposed by Wei et al.\cite{wang2015understanding}, the authors treated fall detection problem as a sequential data processing problem with CSI being the object of interest. While Wei et al. use hidden Markov Model for CARM, Yousefi et al. choose a Long-Short Term Memory (LSTM) network, a common approach to handle sequential data. However, this approach is more of an activity recognition system than merely a fall detection system as they classify 6 different activities, one of which is falling. With this scheme, Yousefi et al. manage to reach an average accuracy of 91\% among all 6 classes.

The most successful attempt to address the robustness issue is FallDeFi by Palipana et al.\cite{palipana2018falldefi}, which employs a feature engineering pipeline to extract robust features for training with simple learning procedures. The method can be factorized into three modules. The first module, named Data Collection and Preprocessing, aims at obtaining clear CSI by using linear interpolation, Discrete Wavelet Transform (DWT) based noised filtering as well as a Principal Component Analysis (PCA) based stream decorrelation and selection algorithm. The second module is the Feature Extraction module. This module receives the stream selection passed from the first module and extracts the features from the stream. Firstly, the Short-Time Fourier Transform (STFT) of the original stream is collected. The authors then use a Power Burst Curve (PBC) to detect events. When an event has been detected, the features are extracted from both the STFT Spectrogram and the PBC. After this, the features are passed to the final module for classification with an SVM classifier. Palipana et al. achieve an accuracy of 80\% when the domains used for training and testing are different.

Recently, domain adaptation method with the adversarial learning paradigm of Jiang et al. \cite{jiang2018towards} can automatically disentangle environment-dependent features, though still rely on spectrogram images \cite{jiang2018towards}. However, this paradigm consists of two limitations. First, adversarial domain adaptation requires us to have data of many domains (a domain is a pair of human and environment). Secondly, domain adaptation is not realistic. For this method to work properly, it must have both data in the source domains (these data are labeled) and the target domains (these data are unlabeled and we are trying to make a prediction in these domains). Therefore, in this research, we are concerned with situations in which we do not have access to the data of the target domain. Shu et al. \cite{shu2018dirt} has also experimented with adapting a vanilla CNN with Virtual Adversarial Domain Adaptation (VADA). However, the performance is not exciting, with only 53.0\% of accuracy for adapted models and 35.7\% for the source-only model for the 6-class dataset released in \cite{yousefi2017survey}. The Decision-boundary Iterative Refinement Training with a Teacher (DIRT-T) in the same paper \cite{shu2018dirt} follows the domain adaptation scheme where the model is refined on the known target domain but does not yield further improvement over VADA.  Based on the work of \cite{domain-generalization-volpi-2018}, we explore adapting the Adversarial Data Augmentation (ADA) method to remove these limitations.

\section{Generalize Wi-Fi-based Fall Detection Models}

\begin{figure}[ht]
    \centering
    \includegraphics[width=\linewidth]{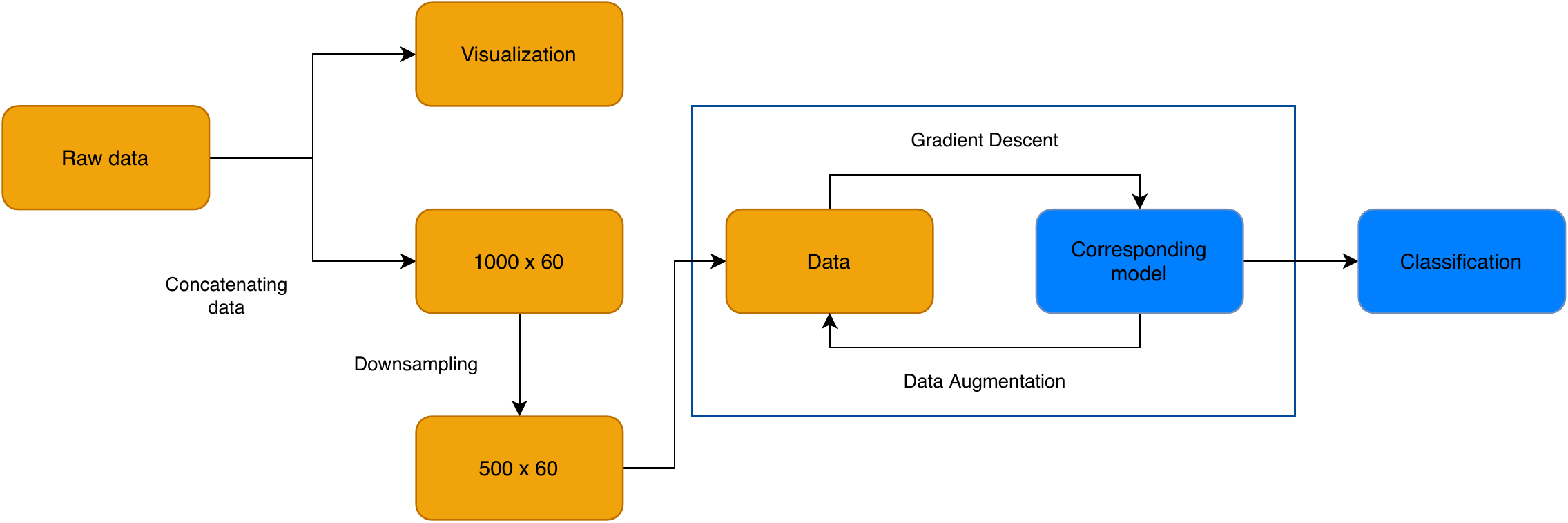}
    \caption{An overview of the procedure}
    \label{fig:model_overview}
\end{figure}

In this work, our primary goal is to experiment applying ADA over suitable network structures for a domain-independent Wi-Fi-based fall detection system. Fig. \ref{fig:model_overview} depicts an overview of our procedure. 

Our methods focus on generalizing across unknown domains beyond the training domains. To do so, we employ the method of Adversarial Data Augmentation introduced by Volpi et al. \cite{domain-generalization-volpi-2018}. The underlying network structure of ADA is tested with various network structures to test their suitability.

\subsection{Data Preprocessing}

\begin{figure}[!htbp]
    \centering
    \includegraphics[width=0.5\textwidth, keepaspectratio]{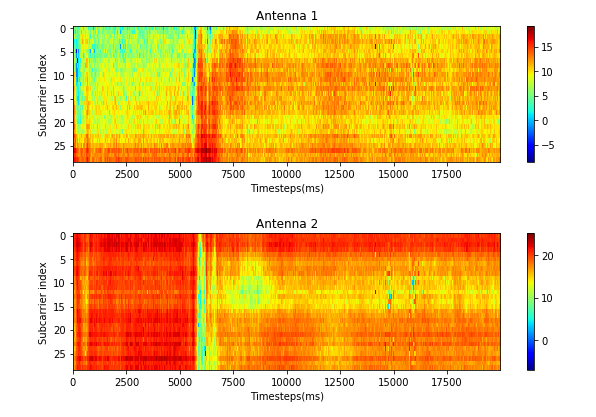}
    \caption{The amplitude heat map from a data sample. Each map correspond to the CSI obtained by an antenna pair.}
    \label{fig:csi_amplitude}
\end{figure}

First, we need to convert the original binary data into numerical matrices. The numerical data is investigated through visualization for better insights. The amplitude heat map of a raw data sample can be found in Fig. \ref{fig:csi_amplitude}. For the sake of information loss tolerance in downsampling, we then concatenate signals from two antennas to create a set of $10000 \times 60$ matrices, each represents the CSI in 10 seconds, matching the source sampling rate of the device at $1000 Hz$. According to the experiments conducted by \cite{wang2015understanding}, there is a correlation between the phase of CSI and human interactions. As we observe in the visualization step, there is information redundancy in overlapped sampling periods. Hence, we can downsample the data to $500 \times 60$ matrices to prevent memory error. 

\subsection{Adversarial Data Augmentation}
\label{ada}
\subsubsection{Overview}
The problem Adversarial Data Augmentation (ADA) \cite{domain-generalization-volpi-2018} tries to address is to get the data from a single source and try to discern other unexplored domains. This problem can be solved by considering the worst-case scenario: The distribution of the target domain is a distance $\rho$ away from the source domain's distribution on the semantic space. The implied distance is the Wasserstein distance \cite{dobrushin1970prescribing}. In other words, Adversarial Data Augmentation aims to solve the following problem:
\begin{equation}
    \min_{\theta \in \Theta} \: \sup_{P: D(P, P_0) \le \rho} \: \mathds{E}[\ell(\theta; (X, Y))]
\end{equation}

In this formula, $\theta \in \Theta$ is our model, while $(X, Y)$ is a source data point, $l$ is the loss function of our choice, and $D(A, B)$ is the distance metric between the two probability distribution $A$, $B$. For this particular paper, the metric we are using is the Wasserstein distance \cite{dobrushin1970prescribing}.
\subsubsection{Wasserstein Distance}
Before going into details, we are going to define a few notations. Let $\theta = (\theta_f, \theta_c)$ be the representation for a model in which $\theta_c$ and $\theta_f$ be the weights of the classification layer and the weights of the rest of the network, respectively.
Let's consider the transportation cost when we move a data point from $\theta_1 = (z, y)$ to $\theta_2 = (z', y')$:
\begin{equation}
    c((z, y),(z', y')) :=\frac{1}{2}\left\|z-z^{\prime}\right\|_{2}^{2}+\infty \cdot 1 \;\; \{y \neq y'\}
\end{equation}

Since we are only interested in cases in which $(z', y')$ is perturbed with respect to the marginal distribution $Z$, $c((z, y), (z', y'))$ is infinity when the label of the two data points are different (meaning $y \neq y'$). Therefore, the transportation cost of data $c_\theta$ with reference to the weight of the final hidden layer, with $g(\theta_f, x)$ be the output of the embedding layers:

\begin{equation}
    c_{\theta}((x, y),(x', y')) :=c((g(\theta_{f} ; x)), y),(g(\theta_{f} ; x'), y'))
\end{equation}

We define the distance for our loss function as follows:
\begin{equation}
    D_{\theta}(P, Q) :=\inf_{M \in \Pi(P, Q)} \mathds{E}_{M}[c_{\theta}((X, Y),(X', Y'))]
\end{equation}

In this notation, $M \in \Pi(P, Q)$ is the set of coupling between two distribution measures, and $inf$ is the infimum of this set. If we can find the solution to the worst-case problem, this solution can ensure a good result in data distributions that are $\rho$ away from the source distributions. The main reason the Wasserstein distance is used is that it allows target distributions $P$ such that $D(P, P_0) \le \rho$ to become realistic representations of the covariate shifts of the source domain while retaining the semantic meaning of the source domain \cite{domain-generalization-volpi-2018}. The space that is actually used is the learned representation since there is a correlation between the distance in the semantic space and the distance in learned representation in high capacity model\cite{DBLP:journals/corr/DosovitskiyB16}.

\subsubsection{Implementation}
Adversarial Data Augmentation \cite{domain-generalization-volpi-2018} is simply running the following procedure for k times\footnote{k can be an arbitrary number. In our implementation, we use k = 100 heuristically.}. The first stage is the \emph{maximization} phase. In this step, new data points are adversarially generated by computing the inner surrogate loss:

\begin{equation}
    \phi_{\gamma}(\theta ;(x_{0}, y_{0})) :=\sup_{x \in \mathcal{X}}\{\ell(\theta ;(x, y_{0}))-\gamma c_{\theta}((x, y_{0}),(x_{0}, y_{0}))\}
\end{equation}

The second stage is the \emph{minimization} phase in which the model is updated using gradient descent using the newly generated data points as well as the original inputs. These two phases are interchanged to continuously create new data points and learn them instantaneously. The generated samples help the model to be more robust against unseen domains because it is generating samples from an unknown domain that are a distance $\rho$ away from the source distributions.

However, it is undetermined what might be the best value for the latent difference  $\rho$. Therefore, we consider different values for the distance  $\rho$ and adversarially generate perturbed samples for each value. After that, we train a model for each value of $\rho$. Consequently, we have an ensemble of models. Because we are not sure which model performs best, during test time, we would choose the most fitting model.

\subsection{Convolutional Neural Network (CNN-ADA)}
\label{cnn}
\begin{figure}[h!]
  \centering
  \includegraphics[width=\linewidth]{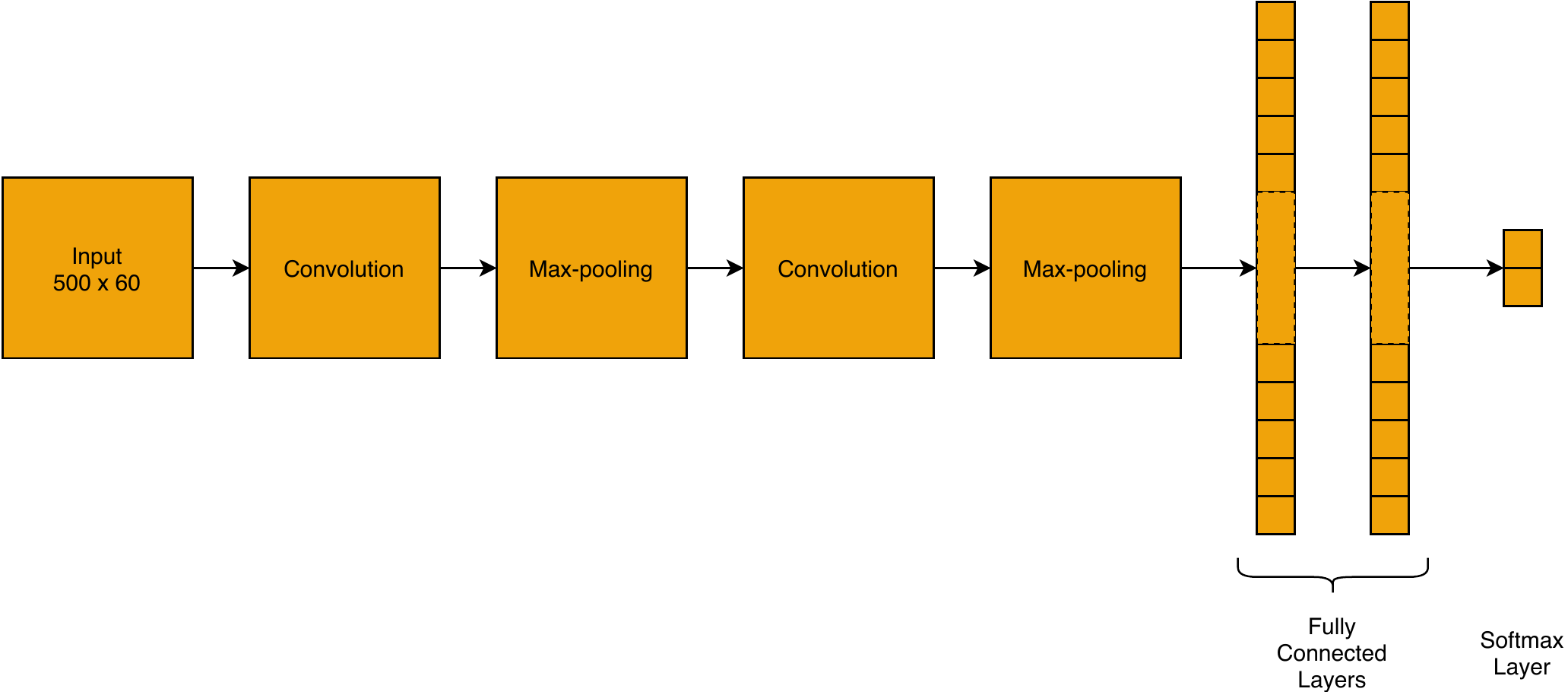}
  \caption{ConvNet architecture}
\end{figure}

Recall that Shu et al. has employed a vanilla $18$-layer CNN for the WiFi activities recognition task \cite{shu2018dirt}. We first evaluate ADA on top of a simple $6$-layer CNN. Firstly, a sample is resized to the shape of $500 \times 60$ to be the input. After that, the input is passed through a convolution layer with $64$ feature maps and a $5 \times 5$ kernel. Subsequently, the tensor goes through a max-pooling layer to reduce the size of the model. In addition to that, there is another convolution layer consisting of $128$ feature maps and a $5 \times 5$ kernel, followed by a supplementary max-pooling layer. Finally, there are two fully connected layers, followed by a softmax layer for classification. 

\subsection{Long-Short Term Memory (LSTM-ADA)}
Although CSI can be evaluated as image data, CSI seems to be more relevant to the category of sequential data \cite{ma2019wifi}. Therefore, we suggest that we adopt a more reasonable approach, such as Recurrent Neural Network (RNN). As RNN is vulnerable to exploding and vanishing gradient, we use Long-Short Term Memory (LSTM) cell for our RNN. First, the original input is put through a series of transpose, reshaping, and splitting operations. After these steps, the resulting tensor $X_1 \in R^{b \times t}$, with $b$ is the number of samples in a batch, $t$ is the number of time steps, is passed to the next component - which is an LSTM cell. For our approach, we are using an LSTM cell with the size of the hidden state vector $h = 200$. The output of the LSTM cell is a tensor $X_2 \in R^{b \times h}$, which is then pushed through a softmax layer of size $h \times c$, with $c$ is the number of classes in the classification task. In our problem, $c = 2$, corresponding to falling and not falling.

\section{Experimentation} 
\subsection{Dataset}
Among two publicly available datasets, Stanford WiFi Activity Recognition \cite{yousefi2017survey} and FallDeFi \cite{palipana2018falldefi}, we employ the FallDeFi dataset as it comes with a clearer description of domains.

The activities in FallDeFi is recorded in 5 environments:
\begin{itemize}
    \item Bathroom and toilet of size $2.9 m \times 4 m$
    \item Two consecutive bedrooms of size $3.1 m \times 5.2 m$
    \item A corridor of size $1.1 m \times 9 m$
    \item A kitchen of size $4.2 m \times 4.5 m$
    \item A lab with a size of $4.6 m \times 7.3 m$
\end{itemize}

Particularly, the transceiver and receiver for recording are two Linux laptops equipped with the Intel 5300 WiFi NIC card and corresponding CSI tool \cite{halperin2011tool}. These laptops are placed with different distances and Line of Sight (LoS) settings for each environment as follow:
\begin{itemize}
    \item Bathroom and toilet: 5m, through two walls with plaster partitions.
    \item Two bedrooms: 5m, through wall
    \item Corridor: 9m, LoS uncluttered
    \item Kitchen: 4m, LoS cluttered
    \item Lab: 7m, non-LoS cluttered
\end{itemize}

Although each laptop is equipped with two external omnidirectional antennas, only the CSI from two antenna pairs (i.e., from 1 transceiver antenna to 2 receiver antennas).

The dataset is collected in two different ways into two subsets A and B. The two subsets are different by the date of collection along with natural chronological changes of involved compartments, the number of actors, and LoS variation created by moving furniture and recording devices. Hence, the dataset can be grouped into 10 domains in total.

\subsection{Experimental result}
\begin{figure}[h!]
\label{cnn_vs_lstm}
  \centering
  \includegraphics[width=\linewidth]{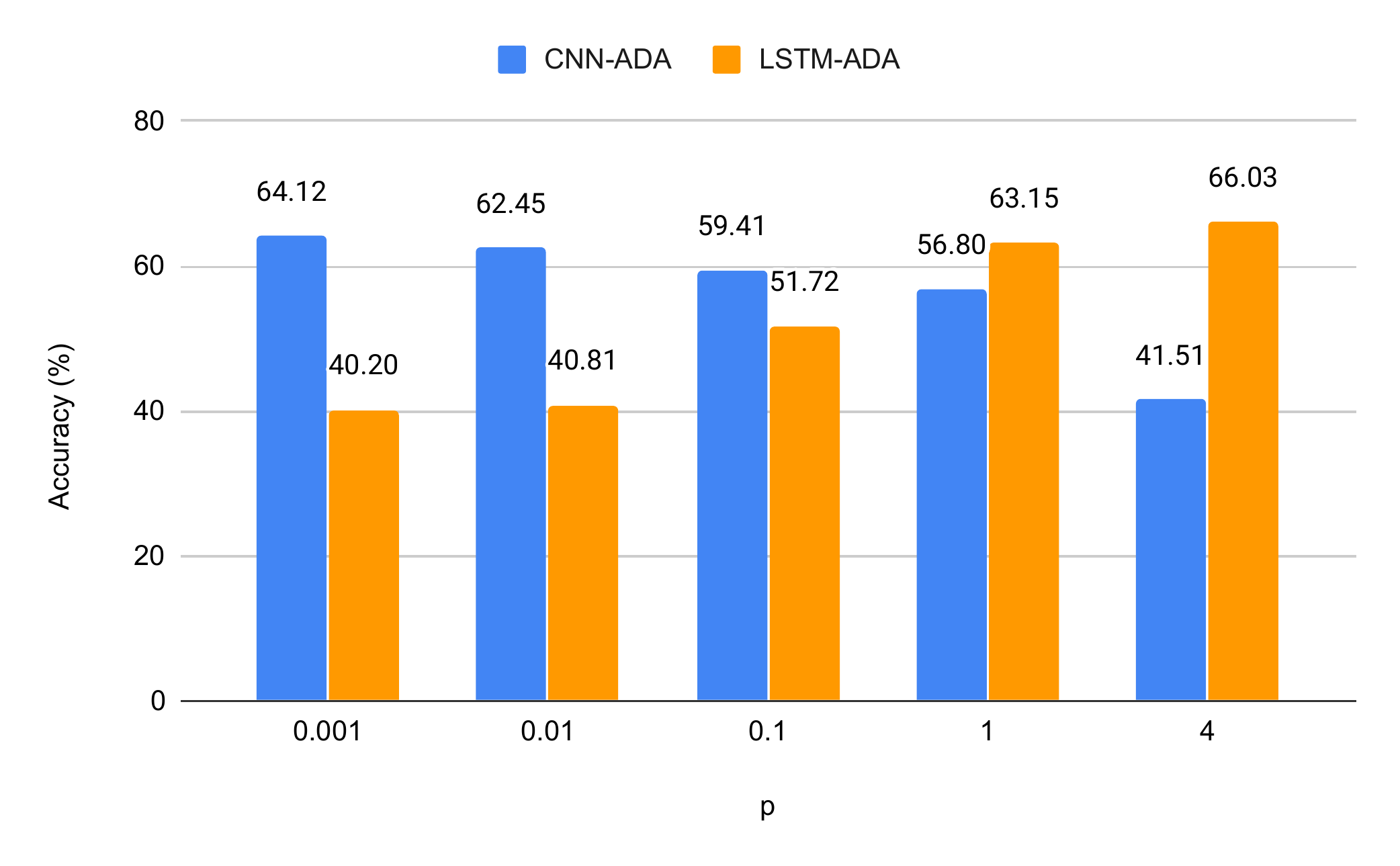}
  \caption{CNN vs LSTM performance among different value of $\rho$}
  \label{fig:ada-results}
\end{figure}

As we are aiming to achieve environment-independence, the experiments are designed to accomplish independence between the training and testing data distributions. Among the $10$ domain employed, we use $9$ domains for the training procedure and $1$ domain for testing the final result. As presented in \ref{ada}, the optimal distance between the source distributions and the distribution of the generated samples $\rho$ is unknown. Therefore, we experiment with different values of $\rho$ to determine which value would give the best performance.

From the evaluation, we got a very thought-provoking result (Fig. \ref{fig:ada-results}). There is a downward trend in the accuracy of the fall detection task with CNN-ADA when we increase the value of $\rho$, and the model achieves the highest accuracy of $64.12\%$ at the $\rho$ value of $0.001$. In contrast, as the value of $\rho$ increases, LSTM-ADA accuracy sees a steady escalation and reaches the highest value of $66.03\%$ at the $\rho$ value of $4$.

Furthermore, we adapt the VADA method of Shu et al. \cite{shu2018dirt} as a binary classification model on FallDeFi. Our VADA adaptation uses the "small" CNN settings described in \cite{shu2018dirt} with an 18-layer vanilla CNN trained using virtual adversarial training. After the training in 10000 iterations on the train set, VADA achieves a $52.7\%$ on the test set (Table \ref{tab:compare-results}).

\begin{table}
\caption{Performance of different models}
\label{tab:compare-results}
\centering
\begin{tabular}{|l|l|}
\hline
\textbf{Model} & \textbf{Accuracy} (\%) \\ \hline
VADA & 52.7\\
\hline
CNN-ADA, $\rho = 0.001$ & 64.12 \\
CNN-ADA, $\rho = 0.01$ & 62.45\\
CNN-ADA, $\rho = 0.1$ & 59.41 \\
CNN-ADA, $\rho = 1$ & 56.60 \\
CNN-ADA, $\rho = 4$ & 41.51\\
\hline
LSTM-ADA, $\rho = 0.001$ & 40.20 \\
LSTM-ADA, $\rho = 0.01$ & 40.81 \\
LSTM-ADA, $\rho = 0.1$ & 51.72 \\
LSTM-ADA, $\rho = 1$ & 63.15 \\
\textbf{LSTM-ADA, $\rho = 4$} & \textbf{66.03} \\
\hline
\end{tabular}
\end{table}


\section{Conclusion and Discussion}
In this paper, we propose applying ADA to improve current Wi-Fi-based fall detection deep learning methods in unseen domains. We conduct experiments with different ADA network structures and confirm that vanilla CNN is not suitable for deep learning-based fall detection systems. Our results with a $6$-layer vanilla CNN reach an accuracy of $64.12\%$, exceeds the results of applying VADA \cite{shu2018dirt} over an $18$-layer vanilla  CNN by $12.12\%$. The results of applying ADA over a simple LSTM also reach an accuracy of $66.03\%$. These results support the postulate that vanilla CNN is not suitable for CSI processing and show contributions of ADA for the generalization task. Unfortunately, this performance is still not good enough for real-world applications, and there is still space for further improvements. Future works should concern evaluating ADA more extensively on other deep networks and incorporating hybrid techniques for CSI processing. State-of-the-art meta-learning methods besides data augmentation are also of concern. 

\section{Acknowledgments}
This research is supported by research funding from the Advanced Program in
Computer Science, University of Science, Vietnam National University - Ho Chi
Minh City.

\bibliographystyle{IEEEtran}
\bibliography{IEEEabrv, mybib}

\begin{thebibliography}{10}
\providecommand{\url}[1]{#1}
\csname url@samestyle\endcsname
\providecommand{\newblock}{\relax}
\providecommand{\bibinfo}[2]{#2}
\providecommand{\BIBentrySTDinterwordspacing}{\spaceskip=0pt\relax}
\providecommand{\BIBentryALTinterwordstretchfactor}{4}
\providecommand{\BIBentryALTinterwordspacing}{\spaceskip=\fontdimen2\font plus
\BIBentryALTinterwordstretchfactor\fontdimen3\font minus
  \fontdimen4\font\relax}
\providecommand{\BIBforeignlanguage}[2]{{%
\expandafter\ifx\csname l@#1\endcsname\relax
\typeout{** WARNING: IEEEtran.bst: No hyphenation pattern has been}%
\typeout{** loaded for the language `#1'. Using the pattern for}%
\typeout{** the default language instead.}%
\else
\language=\csname l@#1\endcsname
\fi
#2}}
\providecommand{\BIBdecl}{\relax}
\BIBdecl

\bibitem{rubenstein2006falls}
L.~Z. Rubenstein, ``Falls in older people: epidemiology, risk factors and
  strategies for prevention,'' \emph{Age and ageing}, vol.~35, no. suppl\_2,
  pp. ii37--ii41, 2006.

\bibitem{friedman2002falls}
S.~M. Friedman, B.~Munoz, S.~K. West, G.~S. Rubin, and L.~P. Fried, ``Falls and
  fear of falling: which comes first? a longitudinal prediction model suggests
  strategies for primary and secondary prevention,'' \emph{Journal of the
  American Geriatrics Society}, vol.~50, no.~8, pp. 1329--1335, 2002.

\bibitem{10.1093/ageing/afm169}
\BIBentryALTinterwordspacing
A.~C. Scheffer, M.~J. Schuurmans, N.~van Dijk, T.~van~der Hooft, and S.~E.
  de~Rooij, ``{Fear of falling: measurement strategy, prevalence, risk factors
  and consequences among older persons},'' \emph{Age and Ageing}, vol.~37,
  no.~1, pp. 19--24, 01 2008. [Online]. Available:
  \url{https://doi.org/10.1093/ageing/afm169}
\BIBentrySTDinterwordspacing

\bibitem{noury2007fall}
N.~Noury, A.~Fleury, P.~Rumeau, A.~Bourke, G.~Laighin, V.~Rialle, and J.~Lundy,
  ``Fall detection-principles and methods,'' in \emph{Engineering in Medicine
  and Biology Society, 2007. EMBS 2007. 29th Annual International Conference of
  the IEEE}.\hskip 1em plus 0.5em minus 0.4em\relax IEEE, 2007, pp. 1663--1666.

\bibitem{yu2008approaches}
X.~Yu, ``Approaches and principles of fall detection for elderly and patient,''
  in \emph{e-health Networking, Applications and Services, 2008. HealthCom
  2008. 10th International Conference on}.\hskip 1em plus 0.5em minus
  0.4em\relax IEEE, 2008, pp. 42--47.

\bibitem{lord1991falls}
C.~J. Lord and D.~P. Colvin, ``Falls in the elderly: Detection and
  assessment,'' in \emph{Engineering in Medicine and Biology Society, 1991.
  Vol. 13: 1991., Proceedings of the Annual International Conference of the
  IEEE}.\hskip 1em plus 0.5em minus 0.4em\relax IEEE, 1991, pp. 1938--1939.

\bibitem{delahoz2014survey}
Y.~S. Delahoz and M.~A. Labrador, ``Survey on fall detection and fall
  prevention using wearable and external sensors,'' \emph{Sensors}, vol.~14,
  no.~10, pp. 19\,806--19\,842, 2014.

\bibitem{rashidi2013survey}
P.~Rashidi and A.~Mihailidis, ``A survey on ambient-assisted living tools for
  older adults,'' \emph{IEEE journal of biomedical and health informatics},
  vol.~17, no.~3, pp. 579--590, 2013.

\bibitem{spasova2014survey}
V.~Spasova and I.~Iliev, ``A survey on automatic fall detection in the context
  of ambient assisted living systems,'' \emph{International journal of advanced
  computer research}, vol.~4, no.~1, p.~94, 2014.

\bibitem{zhang2015survey}
Z.~Zhang, C.~Conly, and V.~Athitsos, ``A survey on vision-based fall
  detection,'' in \emph{Proceedings of the 8th ACM international conference on
  PErvasive technologies related to assistive environments}.\hskip 1em plus
  0.5em minus 0.4em\relax ACM, 2015, p.~46.

\bibitem{ma2019wifi}
Y.~Ma, G.~Zhou, and S.~Wang, ``Wifi sensing with channel state information: A
  survey,'' \emph{ACM Computing Surveys (CSUR)}, vol.~52, no.~3, p.~46, 2019.

\bibitem{wang2019survey}
Z.~Wang, K.~Jiang, Y.~Hou, W.~Dou, C.~Zhang, Z.~Huang, and Y.~Guo, ``A survey
  on human behavior recognition using channel state information,'' \emph{IEEE
  Access}, vol.~7, pp. 155\,986--156\,024, 2019.

\bibitem{domain-generalization-volpi-2018}
R.~Volpi, H.~Namkoong, O.~Sener, J.~Duchi, V.~Murino, and S.~Savarese,
  ``Generalizing to unseen domains via adversarial data augmentation,'' in
  \emph{Proceedings of {NIPS}}, 2018.

\bibitem{palipana2018falldefi}
S.~Palipana, D.~Rojas, P.~Agrawal, and D.~Pesch, ``Falldefi: Ubiquitous fall
  detection using commodity wi-fi devices,'' \emph{Proceedings of the ACM on
  Interactive, Mobile, Wearable and Ubiquitous Technologies}, vol.~1, no.~4, p.
  155, 2018.

\bibitem{halperin2011tool}
D.~Halperin, W.~Hu, A.~Sheth, and D.~Wetherall, ``Tool release: Gathering
  802.11 n traces with channel state information,'' \emph{ACM SIGCOMM Computer
  Communication Review}, vol.~41, no.~1, pp. 53--53, 2011.

\bibitem{Xie:2015:PPD:2789168.2790124}
\BIBentryALTinterwordspacing
Y.~Xie, Z.~Li, and M.~Li, ``Precise power delay profiling with commodity
  wifi,'' in \emph{Proceedings of the 21st Annual International Conference on
  Mobile Computing and Networking}, ser. MobiCom '15.\hskip 1em plus 0.5em
  minus 0.4em\relax New York, NY, USA: ACM, 2015, p. 53–64. [Online].
  Available: \url{http://doi.acm.org/10.1145/2789168.2790124}
\BIBentrySTDinterwordspacing

\bibitem{Wu:2016:WWD:2971648.2971658}
\BIBentryALTinterwordspacing
D.~Wu, D.~Zhang, C.~Xu, Y.~Wang, and H.~Wang, ``Widir: Walking direction
  estimation using wireless signals,'' in \emph{Proceedings of the 2016 ACM
  International Joint Conference on Pervasive and Ubiquitous Computing}, ser.
  UbiComp '16.\hskip 1em plus 0.5em minus 0.4em\relax New York, NY, USA: ACM,
  2016, pp. 351--362. [Online]. Available:
  \url{http://doi.acm.org/10.1145/2971648.2971658}
\BIBentrySTDinterwordspacing

\bibitem{steele2009elderly}
R.~Steele, A.~Lo, C.~Secombe, and Y.~K. Wong, ``Elderly persons’ perception
  and acceptance of using wireless sensor networks to assist healthcare,''
  \emph{International journal of medical informatics}, vol.~78, no.~12, pp.
  788--801, 2009.

\bibitem{wang2017wifall}
Y.~Wang, K.~Wu, and L.~M. Ni, ``Wifall: wangdevice-free fall detection by
  wireless networks,'' \emph{IEEE Transactions on Mobile Computing}, vol.~16,
  no.~2, pp. 581--594, 2017.

\bibitem{zhang2015anti}
D.~Zhang, H.~Wang, Y.~Wang, and J.~Ma, ``Anti-fall: A non-intrusive and
  real-time fall detector leveraging csi from commodity wifi devices,'' in
  \emph{International Conference on Smart Homes and Health Telematics}.\hskip
  1em plus 0.5em minus 0.4em\relax Springer, 2015, pp. 181--193.

\bibitem{wang2017rt}
H.~Wang, D.~Zhang, Y.~Wang, J.~Ma, Y.~Wang, and S.~Li, ``Rt-fall: A real-time
  and contactless fall detection system with commodity wifi devices.''
  \emph{IEEE Trans. Mob. Comput.}, vol.~16, no.~2, pp. 511--526, 2017.

\bibitem{zhang2018wispeed}
F.~Zhang, C.~Chen, B.~Wang, and K.~R. Liu, ``Wispeed: A statistical
  electromagnetic approach for device-free indoor speed estimation,''
  \emph{IEEE Internet of Things Journal}, vol.~5, no.~3, pp. 2163--2177, 2018.

\bibitem{yousefi2017survey}
S.~Yousefi, H.~Narui, S.~Dayal, S.~Ermon, and S.~Valaee, ``A survey on behavior
  recognition using wifi channel state information,'' \emph{IEEE Communications
  Magazine}, vol.~55, no.~10, pp. 98--104, 2017.

\bibitem{wang2015understanding}
W.~Wang, A.~X. Liu, M.~Shahzad, K.~Ling, and S.~Lu, ``Understanding and
  modeling of wifi signal based human activity recognition,'' in
  \emph{Proceedings of the 21st annual international conference on mobile
  computing and networking}.\hskip 1em plus 0.5em minus 0.4em\relax ACM, 2015,
  pp. 65--76.

\bibitem{jiang2018towards}
W.~Jiang, C.~Miao, F.~Ma, S.~Yao, Y.~Wang, Y.~Yuan, H.~Xue, C.~Song, X.~Ma,
  D.~Koutsonikolas \emph{et~al.}, ``Towards environment independent device free
  human activity recognition,'' in \emph{Proceedings of the 24th Annual
  International Conference on Mobile Computing and Networking}.\hskip 1em plus
  0.5em minus 0.4em\relax ACM, 2018, pp. 289--304.

\bibitem{shu2018dirt}
R.~Shu, H.~H. Bui, H.~Narui, and S.~Ermon, ``A dirt-t approach to unsupervised
  domain adaptation,'' \emph{arXiv preprint arXiv:1802.08735}, 2018.

\bibitem{dobrushin1970prescribing}
R.~L. Dobrushin, ``Prescribing a system of random variables by conditional
  distributions,'' \emph{Theory of Probability \& Its Applications}, vol.~15,
  no.~3, pp. 458--486, 1970.

\bibitem{DBLP:journals/corr/DosovitskiyB16}
\BIBentryALTinterwordspacing
A.~Dosovitskiy and T.~Brox, ``Generating images with perceptual similarity
  metrics based on deep networks,'' \emph{CoRR}, vol. abs/1602.02644, 2016.
  [Online]. Available: \url{http://arxiv.org/abs/1602.02644}
\BIBentrySTDinterwordspacing

\end{thebibliography}

\end{document}